\documentclass[conference]{IEEEtran}
\IEEEoverridecommandlockouts

\usepackage{cite}
\usepackage{amsmath,amssymb,amsfonts}
\usepackage{algorithmic}
\usepackage{graphicx}
\usepackage{textcomp}
\usepackage{xcolor}
\usepackage{multirow}
\def\BibTeX{{\rm B\kern-.05em{\sc i\kern-.025em b}\kern-.08em
    T\kern-.1667em\lower.7ex\hbox{E}\kern-.125emX}}
\begin{document}

\title{Resolution-Flexible Decoding for Hybrid \\ Neural Video Representations}

\author{\IEEEauthorblockN{Taiga Hayami}
\IEEEauthorblockA{\textit{Graduate School of FSE,} \\
\textit{Waseda University}\\
Tokyo, Japan \\
hayatai17@fuji.waseda.jp}
\and
\IEEEauthorblockN{Masaya Takabe}
\IEEEauthorblockA{\textit{Graduate School of FSE,} \\
\textit{Waseda University}\\
Tokyo, Japan \\
masaya.ta@asagi.waseda.jp}
\and
\IEEEauthorblockN{Hiroshi Watanabe}
\IEEEauthorblockA{\textit{Graduate School of FSE,} \\
\textit{Waseda University}\\
Tokyo, Japan \\
hiroshi.watanabe@waseda.jp}
}

\maketitle

\begin{abstract}
Neural video representations (NVRs) represent videos using neural network parameters and, in hybrid formulations, frame-wise latent embeddings.
In conventional hybrid NVRs, the spatial size of frame-wise latent embeddings and the decoder upsampling schedule depend on the target resolution, potentially requiring large and non-uniform upsampling factors.
In this paper, we propose a resolution-flexible decoder framework for hybrid NVRs.
The decoder is constructed from uniform \(2\times\) upsampling stages, whose target feature sizes are obtained by tracing the spatial resolution backward from the final output resolution.
After each upsampling stage, the feature map is aligned with the target size by minimal padding or cropping when necessary.
To support this progressive decoding process, we further use intermediate reconstruction supervision and a reconstruction-difficulty-aware frame sampling strategy based on recent frame-wise losses.
The framework preserves the basic representation format of hybrid NVRs and can therefore be applied to different backbones.
Experiments on the UVG dataset at resolutions of \(1080\times1920\) and \(810\times1440\) show that the proposed approach improves reconstruction quality over the corresponding NVR baselines.

\end{abstract}


\section{Introduction}
Neural video representations (NVRs), which build on implicit neural representations (INRs) \cite{siren}, have recently attracted attention as a video-specific representation and compression paradigm.
These methods \cite{nerv, enerv, hnerv, dnerv, ffnerv, hinerv, pnerv, boost, dsnerv, tree} model a video with a neural network that takes either a frame index or a frame-wise latent embedding as input and reconstructs the corresponding RGB frame.
Under this formulation, video coding can be viewed as the optimization of a video-specific neural network.
Compression is then achieved by quantizing and entropy coding the trained network weights and, when applicable, the latent embeddings, while decoding requires only a forward pass through the trained model.

Index-based NVRs \cite{nerv, enerv} generate each frame directly from its frame index.
Hybrid NVRs \cite{hnerv, dnerv}, in contrast, improve representation capacity by introducing content-adaptive latent embeddings as decoder inputs.
In this setting, the latent embeddings serve as explicit frame-wise representations, while the decoder acts as an implicit representation shared across all frames.
However, hybrid NVRs still require the spatial size of the latent embeddings and the decoder upsampling schedule to be compatible with the output-frame resolution.
These quantities are commonly chosen so that the product of the intermediate feature size and the stage-wise upsampling factors exactly matches the target resolution.
Prior studies \cite{hnerv, dsnerv} also report that adjusting video resolutions to sizes such as \(960\times1920\) and \(480\times960\), while maintaining an aspect ratio of \(1:2\), contributes to improved model performance.

\begin{figure}[tb]
\centerline{\includegraphics[width=0.8\columnwidth]{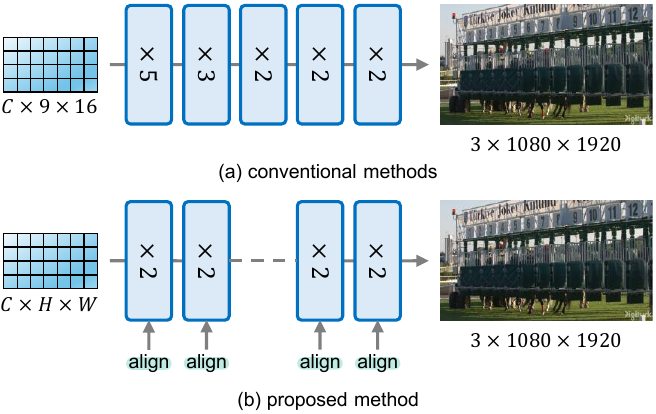}}
\caption{
Comparison of conventional and proposed decoder designs.
Conventional hybrid NVRs use resolution-dependent, non-uniform upsampling schedules, whereas the proposed method reconstructs the target frame using uniform \(2\times\) upsampling stages followed by spatial alignment.
}
\label{comp}
\end{figure}

For example, when reconstructing FHD video at a resolution of \(1080 \times 1920\) (height \(\times\) width), conventional designs require large and non-uniform upsampling factors, such as \(5\times\) and \(3\times\), as shown in Fig. \ref{comp}.
Consequently, spatial resolution conversion is concentrated in specific decoder stages.
Moreover, because the embedding grid is constrained by the factorization of the final resolution, this setting requires at least a \(9 \times 16\) spatial grid for the latent embeddings.
In a typical hybrid NVR configuration with 16 embedding channels, a 600-frame UVG sequence \cite{uvg} therefore requires approximately 1.38M embedding parameters.
Under a fixed total parameter budget, this embedding allocation reduces the number of parameters available for the decoder, which may limit representation capacity, particularly in low-bitrate or compact-model settings.

\begin{figure*}[tb]
\centerline{\includegraphics[width=1.8\columnwidth]{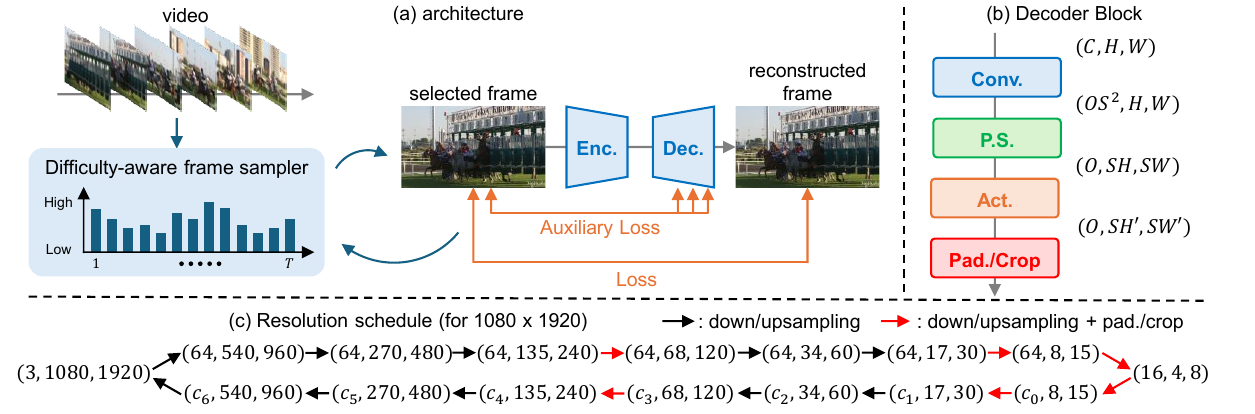}}
\caption{
Overview of the proposed framework.
(a) Overall architecture consisting of a difficulty-aware frame sampler, an encoder, and a decoder.
(b) Proposed decoder block, which performs convolution, pixel shuffle, and activation, followed by spatial alignment through padding or cropping when necessary.
(c) Example resolution schedule for reconstructing a \(1080 \times 1920\) frame.
}
\label{all}
\end{figure*}

In this paper, we propose a hybrid NVR framework in which all decoder stages are constructed using uniform \(2\times\) upsampling.
In the proposed method, the sequence of target resolutions is determined by tracing the spatial resolution backward from the final frame resolution.
At each decoder stage, \(2\times\) upsampling is first applied, and the resulting feature map is then aligned with the target resolution using minimal padding or cropping.
This design avoids relying on large upsampling factors determined by the final resolution and allows the decoder to follow a uniform and progressive reconstruction path.
To supervise reconstruction at multiple spatial scales, we apply intermediate reconstruction losses to selected decoder stages.

We further introduce a reconstruction-difficulty-aware frame sampling strategy.
Instead of sampling frames uniformly throughout training, the proposed sampler estimates frame-wise difficulty scores from recent reconstruction losses and samples frames from a mixture of difficulty-based and uniform distributions.
This strategy allocates more training opportunities to difficult frames while maintaining coverage of the entire video.

\section{Related Work}
\subsection{Neural Video Representation}
Neural video representations encode a video in neural network parameters and reconstruct each frame through a forward pass.
NeRV \cite{nerv} first introduced a frame-wise neural representation that maps a frame index to the corresponding RGB frame, enabling video compression through quantization, pruning, and entropy coding of the learned network weights.
HNeRV \cite{hnerv} further improved reconstruction quality by introducing content-adaptive frame-wise embeddings extracted by a learnable encoder.
Subsequent methods have enhanced NVRs through improved temporal modeling or multi-scale feature design.
FFNeRV \cite{ffnerv} exploits temporal redundancy using flow-guided frame-wise representations and temporal grids, HiNeRV \cite{hinerv} improves compression through hierarchical positional encoding, and PNeRV \cite{pnerv} introduces a pyramidal representation to improve spatial consistency.
HNeRV-Boost \cite{boost} further improves multiple NVR backbones by using a conditional decoder to align intermediate features with target frames.

In contrast to these methods, our work focuses on the resolution dependency of hybrid NVR decoders.
In conventional hybrid-NVR designs, both the latent embedding size and the decoder upsampling schedule are constrained by the factorization of the target resolution, often resulting in large and non-uniform upsampling factors.
We address this issue by constructing the decoder with uniform \(2\times\) upsampling stages and lightweight spatial alignment, thereby providing a more flexible and progressively structured reconstruction path.

\subsection{Sampling Strategies}
Most NVR methods train video representations by sampling frames uniformly from the target video.
Although uniform sampling is simple and widely adopted, it treats all frames equally even though reconstruction difficulty can vary over time because of motion, texture complexity, occlusion, and scene changes.
This property can make optimization less efficient, especially under limited model capacity or short training schedules.

Adaptive sampling has been studied as a way to allocate training effort to informative or difficult samples.
In the context of neural fields, Kheradmand \textit{et al.} \cite{mining} proposed Soft Mining, which accelerates training by selecting samples according to an importance distribution derived from training errors.
In neural video representation, Tree-NeRV \cite{tree} introduces a tree-structured feature grid with non-uniform temporal sampling, assigning denser feature nodes to temporal regions with high reconstruction errors.

\section{Proposed Method}

\subsection{Overall Framework}

Let \(\mathbf{I}_t \in \mathbb{R}^{3 \times H \times W}\) denote the \(t\)-th frame of a target video.
In hybrid NVRs, each frame is represented by a content-adaptive latent embedding \(\mathbf{z}_t\), and a shared decoder \(D_\theta\) reconstructs the corresponding RGB frame as \(\hat{\mathbf{I}}_t = D_\theta(\mathbf{z}_t)\).
During training, an encoder \cite{convnext} extracts a latent embedding from each input frame.
After training, the video is represented by the trained decoder parameters and the frame-wise latent embeddings.

The proposed method aims to reduce the dependence of decoder design on the factorization of the target resolution.
Instead of using resolution-dependent and non-uniform upsampling factors, the decoder is built from repeated \(2\times\) upsampling stages.
Given the output resolution \((H, W)\) and the number of decoder stages \(L\), we determine the target spatial size of each decoder stage by tracing backward from the final resolution:
\begin{equation}
    (H_l,W_l) = \operatorname{round}_{\mathrm{even}}\left( \frac{(H,W)}{2^{L-l}} \right), \quad l=0,\ldots,L,
\end{equation}
where \(\operatorname{round}_{\mathrm{even}}(\cdot)\) denotes round-to-even rounding.
Here, \((H_L, W_L) = (H, W)\), and \((H_0, W_0)\) denotes the initial spatial size of the decoder feature map.
The rounding is applied component-wise. Thus, the initial grid can be selected without requiring that every output dimension be divisible by \(2^L\), while the subsequent alignment resolves the rounding discrepancies.

The proposed framework consists of three components: a uniform \(2\times\) decoder with spatial alignment, intermediate reconstruction supervision, and reconstruction-difficulty-aware frame sampling.
The decoder design and intermediate supervision facilitate optimization of the progressive reconstruction process, while the sampling strategy allocates more training opportunities to frames that are difficult to reconstruct.
Because the framework does not alter the basic representation format, it can be applied to different hybrid NVR backbones, such as HNeRV and HNeRV-Boost.

\subsection{Decoder Block}

Let \(\mathbf{F}_l \in \mathbb{R}^{C_l \times H_l \times W_l}\) denote the input feature map of the \(l\)-th decoder stage.
As shown in Fig. \ref{all}, each decoder block first performs \(2\times\) upsampling using convolution and pixel shuffle\cite{ps}.
The upsampled feature map is then aligned to the predefined target size when necessary:
\begin{equation}
\mathbf{F}_{l+1} = A_{H_{l+1},W_{l+1}}\!\left(\phi_l(\mathbf{F}_l)\right),
\end{equation}
where \(\phi_l(\cdot)\) denotes the \(2\times\) upsampling operation composed of convolution, pixel shuffle, and activation.
The function \(A_{H_{l+1},W_{l+1}}(\cdot)\) aligns the feature map with the target resolution \((H_{l+1}, W_{l+1})\).
Specifically, it applies center cropping when the feature map is larger than the target size and padding when it is smaller.
If the feature size already matches the target resolution, the alignment operation reduces to the identity mapping.
By repeating this block, the decoder forms a uniform progressive reconstruction path from the latent feature map to the output frame.
Unlike conventional HNeRV-style designs that rely on large factors such as \(5\times\) or \(3\times\), the proposed decoder distributes spatial resolution conversion across multiple \(2\times\) stages.

\subsection{Loss Function}

The final frame is reconstructed from the last decoder feature map, while auxiliary reconstruction heads are attached to selected intermediate stages to stabilize progressive \(2\times\) decoding.
For each supervised stage \(l \in \mathcal{S}\), the auxiliary head predicts a low-resolution RGB image, which is supervised by the target frame resized to the same spatial resolution.
The training loss for frame \(t\) is defined as
\begin{equation}
    \mathcal{L}(t) = \ell(\hat{\mathbf{I}}_t,\mathbf{I}_t) + \sum_{l \in \mathcal{S}} \lambda_l \ell \left( h_l(\mathbf{F}_l), R_l(\mathbf{I}_t) \right),
\end{equation}
where \(\ell(\cdot)\) denotes the reconstruction loss, \(h_l(\cdot)\) is the auxiliary head, \(R_l(\cdot)\) resizes the target frame to the resolution of stage \(l\), and \(\lambda_l\) is the auxiliary-loss weight.
The auxiliary heads are used only during training.

\begin{table}[t]
\centering
\caption{Comparison of reconstruction quality on the UVG dataset in PSNR (3.0M, 1080$\times$1920 and 810$\times$1440).}
\label{uvg}
{\fontsize{8}{9}\selectfont
\setlength{\tabcolsep}{1.2pt}
\renewcommand{\arraystretch}{1.0}
\begin{tabular*}{\columnwidth}{@{\extracolsep{\fill}}lrrrrrrrr@{}}
\hline
Method & Beauty & Bosph & Honey & Jockey & Ready & Shake & Yacht & Avg. \\
\hline
\multicolumn{9}{c}{\textit{1080 $\times$ 1920}} \\
NeRV       & 33.06 & 33.43 & 37.13 & 31.55 & 24.97 & 32.96 & 28.01 & 31.59 \\
\hline
HNeRV      & 33.34 & 31.55 & 39.15 & 32.26 & 25.52 & 33.95 & 28.32 & 32.01 \\
HNeRV+Ours & \textbf{33.70} & \textbf{35.82} & \textbf{39.24} & \textbf{33.05} & \textbf{27.46} & \textbf{34.98} & \textbf{29.47} & \textbf{33.39} \\
\hline
Boost      & 33.79 & 36.13 & \textbf{39.64} & 34.21 & 28.09 & \textbf{35.87} & 29.27 & 33.86 \\
Boost+Ours & \textbf{33.85} & \textbf{36.64} & 39.63 & \textbf{34.81} & \textbf{28.90} & 35.73 & \textbf{29.57} & \textbf{34.16} \\
\hline
\multicolumn{9}{c}{\textit{810 $\times$ 1440}} \\
NeRV       & 33.89 & 33.23 & 37.80 & 32.28 & 25.63 & 33.59 & 29.50 & 32.27 \\
\hline
HNeRV      & 33.96 & 32.97 & 39.14 & 32.56 & 25.49 & 34.00 & 29.43 & 32.51 \\
HNeRV+Ours & \textbf{34.26} & \textbf{35.60} & \textbf{39.27} & \textbf{33.54} & \textbf{28.01} & \textbf{34.99} & \textbf{30.69} & \textbf{33.77} \\
\hline
Boost      & 34.40 & 35.81 & 39.56 & 34.80 & 28.63 & \textbf{35.85} & 30.82 & 34.27 \\
Boost+Ours & \textbf{34.47} & \textbf{36.79} & \textbf{39.60} & \textbf{35.44} & \textbf{30.04} & 35.77 & \textbf{31.39} & \textbf{34.79} \\
\hline
\end{tabular*}
}
\end{table}

\subsection{Reconstruction-Difficulty-Aware Sampling}

In standard NVR training, frames are typically sampled uniformly from the target video.
However, reconstruction difficulty varies across frames because of motion, texture complexity, occlusion, and scene changes.
Uniform sampling may therefore allocate insufficient training iterations to difficult frames, especially in compact-model settings.

Unlike pixel- or ray-level neural field sampling \cite{mining,fast}, our method operates on the finite set of video frames.
Accordingly, rather than constructing dense spatial error maps, we maintain a scalar difficulty score for each frame and update it using recent reconstruction losses.
Specifically, the frame-wise difficulty score \(q_t\) is updated through an exponential moving average:
\begin{equation}
    q_t \leftarrow \beta q_t + (1-\beta)d_t,
\end{equation}
where \(d_t\) is the reconstruction loss observed for frame \(t\) and \(\beta\) is the momentum coefficient.
The score \(q_t\) is updated whenever frame \(t\) is sampled.

Frames are then sampled from a mixture of difficulty-based and uniform distributions:
\begin{equation}
    p_t = (1-\rho) \frac{q_t}{\sum_{j=1}^{N} q_j} + \rho \frac{1}{N},
\end{equation}
where \(N\) is the number of training frames and \(\rho\) controls the uniform mixing ratio.
The uniform component ensures that all frames retain a nonzero sampling probability.
The sampled loss is reweighted as
\begin{equation}
    \mathcal{L}_{\mathrm{sampled}}(t) = (N p_t)^{-\alpha}\mathcal{L}(t),
\end{equation}
where \(\alpha \in [0,1]\) controls the strength of the correction.

\section{Experiment}

\begin{figure}[tb]
\centerline{\includegraphics[width=1.0\columnwidth]{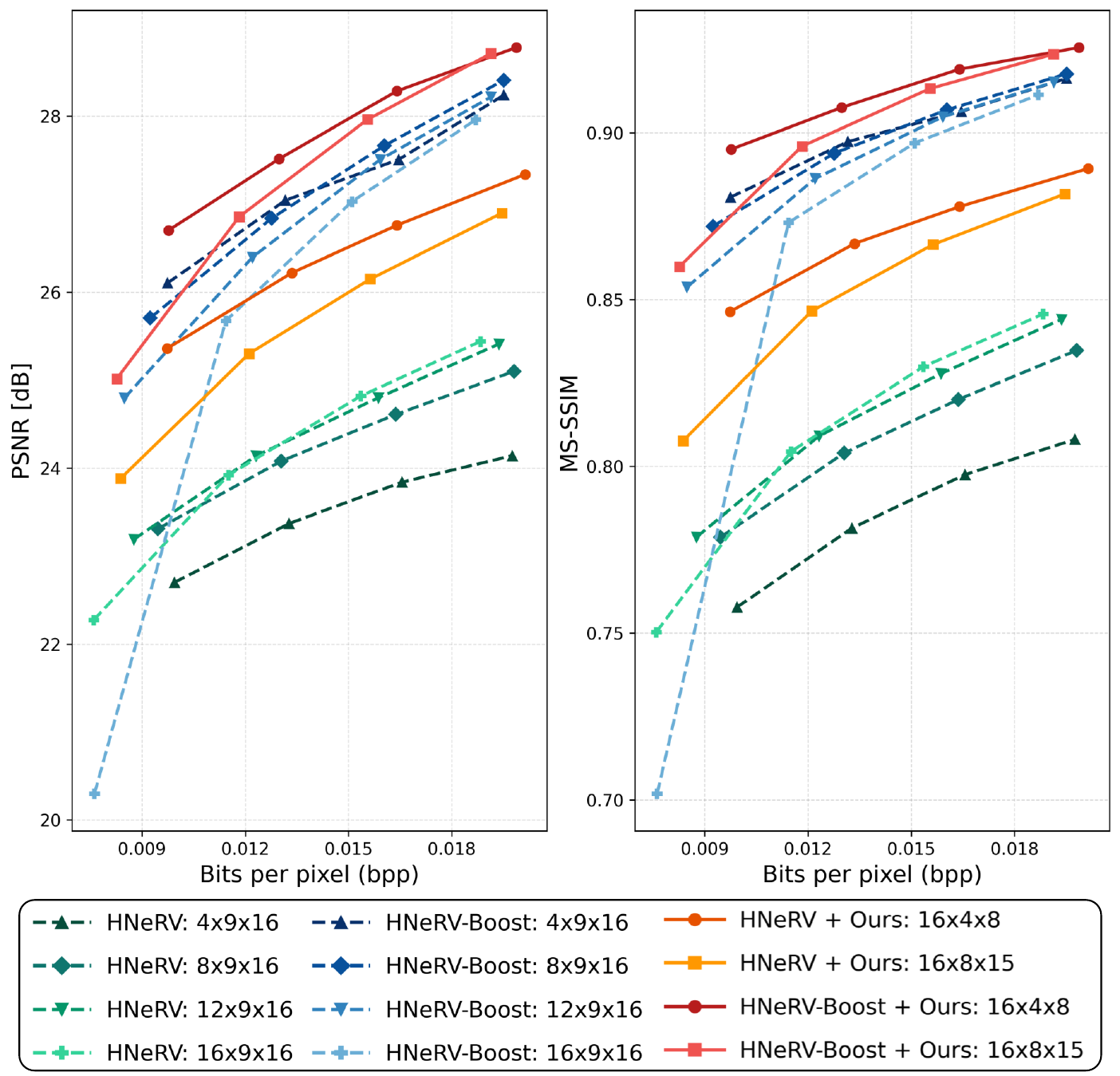}}
\caption{
Rate--distortion curves on the ``ReadySetGo'' sequence at \(1080\times1920\).
}
\label{rd}
\end{figure}

\subsection{Experimental Setup}

We evaluate seven video sequences from the UVG dataset \cite{uvg}.
To assess resolution flexibility, we conduct experiments at two resolutions: \(1080 \times 1920\) and \(810 \times 1440\).
Following the standard NVR setting, each method is optimized independently for each target video.
Reconstruction quality is evaluated using PSNR and MS-SSIM.

We use HNeRV \cite{hnerv} and HNeRV-Boost \cite{boost} as hybrid NVR baselines, and denote the proposed framework applied to each baseline as ``HNeRV + Ours'' and ``HNeRV-Boost + Ours'', respectively.
At \(1080 \times 1920\), the latent embedding sizes are \(16 \times 4 \times 8\) and \(16 \times 8 \times 15\) for HNeRV + Ours and HNeRV-Boost + Ours, respectively.
At \(810 \times 1440\), the corresponding sizes are \(16 \times 3 \times 6\) and \(16 \times 6 \times 11\), where the dimensions represent channels, height, and width.
Intermediate reconstruction supervision is applied to the last \(s\) intermediate decoder stages, with \(s=4\) for HNeRV + Ours and \(s=2\) for HNeRV-Boost + Ours.
The auxiliary-loss weights \(\lambda_l\) are set to \((0.01, 0.025, 0.05, 0.1)\) and \((0.025, 0.05)\), respectively.
The low-resolution targets \(R_l(\mathbf{I}_t)\) are generated by region-based resizing.
For reconstruction-difficulty-aware sampling, the frame-wise difficulty scores are updated with \(\beta = 0.9\), and the sampling probabilities are computed with a uniform mixing ratio of \(\rho = 0.1\).
The sampled loss is reweighted using \(\alpha = 0.6\).
Uniform sampling is applied during the first and last 30 epochs as warm-up and cool-down periods, respectively.

\subsection{Results}

The reconstruction quality on the UVG dataset at both resolutions is summarized in Table \ref{uvg}.
At \(1080 \times 1920\) and \(810 \times 1440\), HNeRV + Ours improves the average PSNR by 1.38 and 1.26 dB over HNeRV, respectively, with improvements on all seven sequences.
HNeRV-Boost + Ours achieves average gains of 0.30 and 0.52 dB over HNeRV-Boost, improving five and six sequences, respectively, with slight decreases on HoneyBee and ShakeNDry at the former resolution and ShakeNDry at the latter.
These results indicate improvements in average reconstruction quality at both evaluated resolutions, although the gains vary across sequences.

We show the rate--distortion (RD) curves of the compressed models with different latent embedding sizes at \(1080 \times 1920\) in Fig. \ref{rd}.
HNeRV + Ours outperforms HNeRV in both PSNR and MS-SSIM, while HNeRV-Boost + Ours generally achieves higher reconstruction quality than HNeRV-Boost at comparable bitrates.
Notably, HNeRV-Boost exhibits substantial quality degradation at low bitrates when larger latent embeddings are used, whereas the proposed framework maintains relatively high reconstruction quality.

\begin{figure}[tb]
\centerline{\includegraphics[width=\columnwidth]{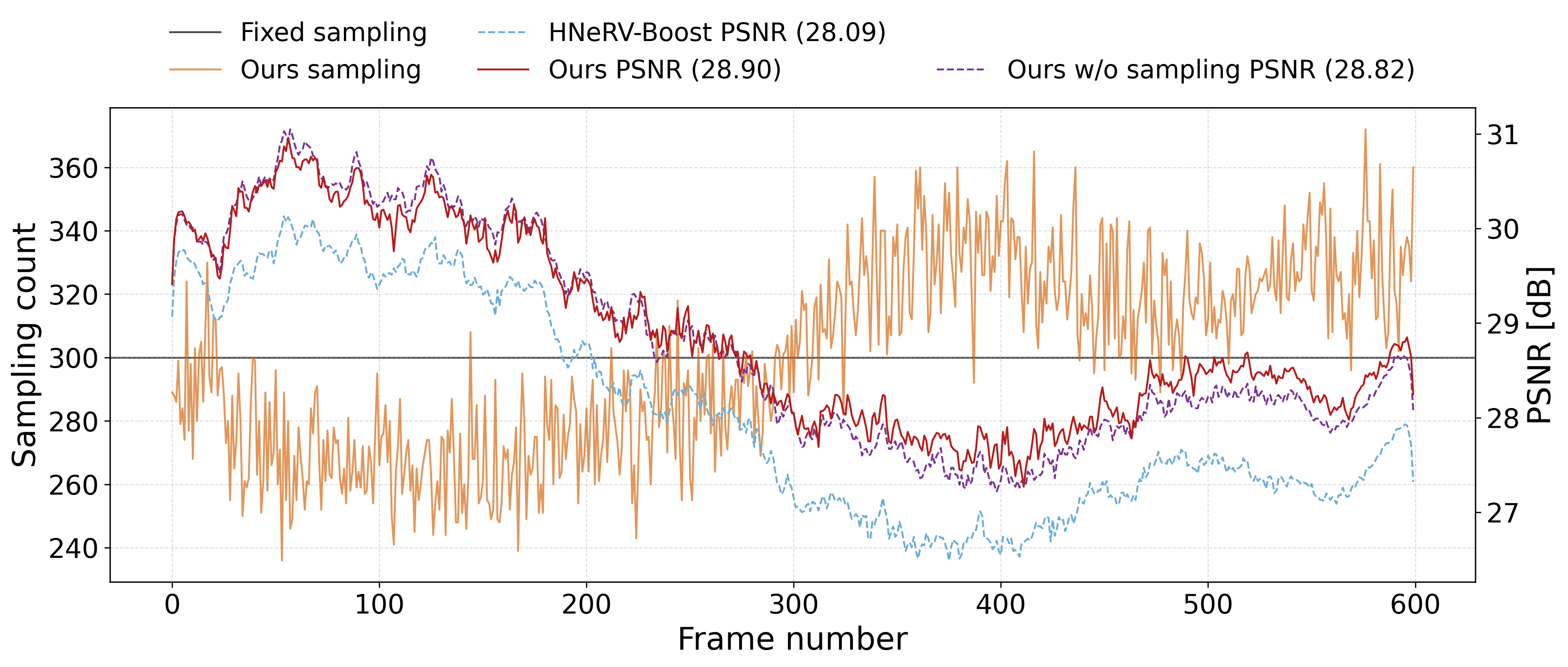}}
\caption{
Frame-wise sampling behavior on the ``ReadySetGo'' sequence at \(1080\times1920\).
The left axis shows the number of samples assigned to each frame, and the right axis shows the frame-wise PSNR.
``Ours w/o sampling'' denotes the proposed decoder trained with uniform frame sampling.
The values in parentheses in the legend indicate the average PSNR over all frames.
}
\label{sampling}
\end{figure}

Fig. \ref{sampling} compares the frame-wise sampling frequency and reconstruction PSNR on ReadySetGo at \(1080 \times 1920\).
Here, ``Ours w/o sampling'' denotes the proposed framework trained with uniform frame sampling.
Compared with uniform sampling, the proposed strategy allocates more samples to frames with relatively low PSNR.
This result suggests that difficulty-aware sampling provides an additional, albeit modest, improvement in reconstruction quality for the evaluated sequence.

\section{Conclusion}

In this paper, we proposed a resolution-flexible decoder framework for hybrid neural video representations.
The proposed method constructs the decoder using uniform \(2\times\) upsampling stages and aligns intermediate feature maps with their target spatial dimensions through padding or cropping when necessary.
We also introduced intermediate reconstruction supervision and reconstruction-difficulty-aware frame sampling to stabilize optimization and allocate more training opportunities to difficult frames.
Experiments on the UVG dataset show that the proposed framework improves reconstruction quality over the corresponding NVR baselines while preserving the basic representation format of hybrid NVRs.

\vspace{12pt}
\end{document}